\documentclass[journal]{IEEEtran}
\ifCLASSINFOpdf
\else
\fi
\usepackage{amssymb}
\usepackage{fancyhdr}
\usepackage{graphicx} 
\usepackage{epsfig}
\usepackage{multirow}
\usepackage[center]{caption}
\usepackage{subcaption}
\usepackage{setspace}
\usepackage{amsmath}
\usepackage{float}
\usepackage{subfig}
\usepackage{cite}
\usepackage{array}
\usepackage{verbatim}
\usepackage{graphicx} 
\usepackage{amsfonts}
\usepackage{MnSymbol}%
\twocolumn
\date{}
\begin{document}
\title{Exact Least Squares Algorithm for Signal Matched Synthesis Filter
Bank:Part II}
\author{Binish~Fatimah and S.~D.~Joshi~
\thanks{The authors are with the Department
of Electrical  Engineering, Indian Institute of Technology, New Delhi 110016 India e-mail: binish.fatimah@ee.iitd.ac.in., sdjoshi@ee.iitd.ac.in}}



\maketitle
\begin{abstract}

In the companion paper, we proposed a concept of signal matched whitening filter bank and developed a time and order recursive, fast least squares algorithm for the same. Objective of part II of the paper is two fold: first is to define a concept of signal matched synthesis filter bank, hence combining definitions of part I and part II we obtain a filter bank matched to a given signal. We also develop a fast time and order recursive, least squares algorithm for obtaining the same. The synthesis filters, obtained here, reconstruct the given signal only and not every signal from the finite energy signal space (i.e. belonging to $L^2(\mathbb{R})$), as is usually done. The recursions, so obtained, result in a lattice-like structure. Since the filter parameters are not directly available, we also present an order recursive algorithm for the computation of signal matched synthesis filter bank coefficients from the lattice parameters. The second objective is to explore the possibility of using synthesis side for modeling of a given stochastic process. Simulation results have also been presented to validate the theory.

\end{abstract}


\section{Introduction}


In the companion paper \cite{PartI} we have presented a concept of signal matched whitening filter bank. The objective, of part II of the series, is two-fold: first is to define a concept of  signal matched synthesis filter bank corresponding to the analysis filter bank obtained in part I \cite{PartI}. It can be easily seen that synthesis filter can be found in, at least two different ways, one is to use the theory of perfect reconstruction filter bank, for the analysis filter bank obtained in the accompanying paper. In that case the complete filter bank, i.e. the analysis side of part I followed by the synthesis side, so obtained, strictly speaking should not be called a ``signal matched filter bank". The second approach, as opted in this work, is to find a synthesis filter bank that reconstructs the given signal only and not every signal belonging to $L^2(\mathbb{R})$, we term this as ``signal matched synthesis filter bank", a fact which was noted in \cite{Nalbalwar_phd}. Therefore, part I and part II, provide a filter bank that is matched to the given signal. The second objective of the paper is to explore the possibility of using the synthesis filter bank as a modeling strategy for a given stochastic process. It should be noted that 
the class of  signals that can be generated using this model, can  be whitened using the analysis filter bank of \cite{PartI}, with an optimized coding gain.

Modeling of stochastic processes, in a multiscale framework, has been an active area of research for more than two decades. In \cite{wilsky_modeling} Basseville et. al. proposed a  multiresolution modeling, by indexing stochastic process by nodes on lattice/tree. Different depths, on lattice/tree, correspond to the different scales of the model, representing signal at different resolution. It is also shown in \cite{wilsky_modeling} that wavelets are suited for the proposed modeling scenario.

 In the deterministic scenario, a multiresolution signal approximation, signal being any element of $L^{2}(\mathbb{R}^{n})$, was proposed by Mallat \cite{Mallat_wavelet}. Here, the signal is represented in terms of diadic dilates and translates of a single function $\psi(\mathbf{x})$ called mother wavelet. Wavelets as orthonormal bases obtained here, are found to be very important for a variety of signal modeling applications. Regarding the coefficients in the wavelet expansion as zero-mean, independent random variables, with an associated special statistical properties, Wornell \cite{Wornell_fractal} showed that time averaged PSD of such a process exhibits generalized $1/f^{\beta}$ character. The Discrete Wavelet Transform (DWT)  decomposes a signal into frequency channels with equal bandwidth on a logarithmic scale. The multirate filter bank, on the other hand, have no such restriction and thus provide greater flexibility than DWT.

Vrcelj et. al. \cite{vrcelj} proposed a least squares signal approximation, for any arbitrary square summable signal, using a multirate signal model. The model parameters, i.e. the synthesis filters, are assumed to be known and the analysis filters are obtained using the perfect reconstruction filter bank theory.
The synthesis filter bank inputs, required for the signal representation, are obtained from the corresponding analysis side.
 The model discussed in \cite{vrcelj} is restricted by the fixed synthesis filters. In \cite{Lall_modelling} Lall et. al. proposed a multiscale modeling technique employing a sub-band coder. They proposed a model, for cyclostationary processes, in which each stage, of the synthesis side of filter bank, models the signal as it evolves across scale.
 The computation of model parameters, however, required an extrinsic concept of blurring filter. Nalbalwar et. al. \cite{Nalbalwar_NCC}, proposed a signal matched Non-uniform synthesis filter bank for, modeling of stochastic process. 
 
\begin{figure}[H]
\hspace{-1cm}
\includegraphics[width=0.6\textwidth, height=0.4\textwidth]{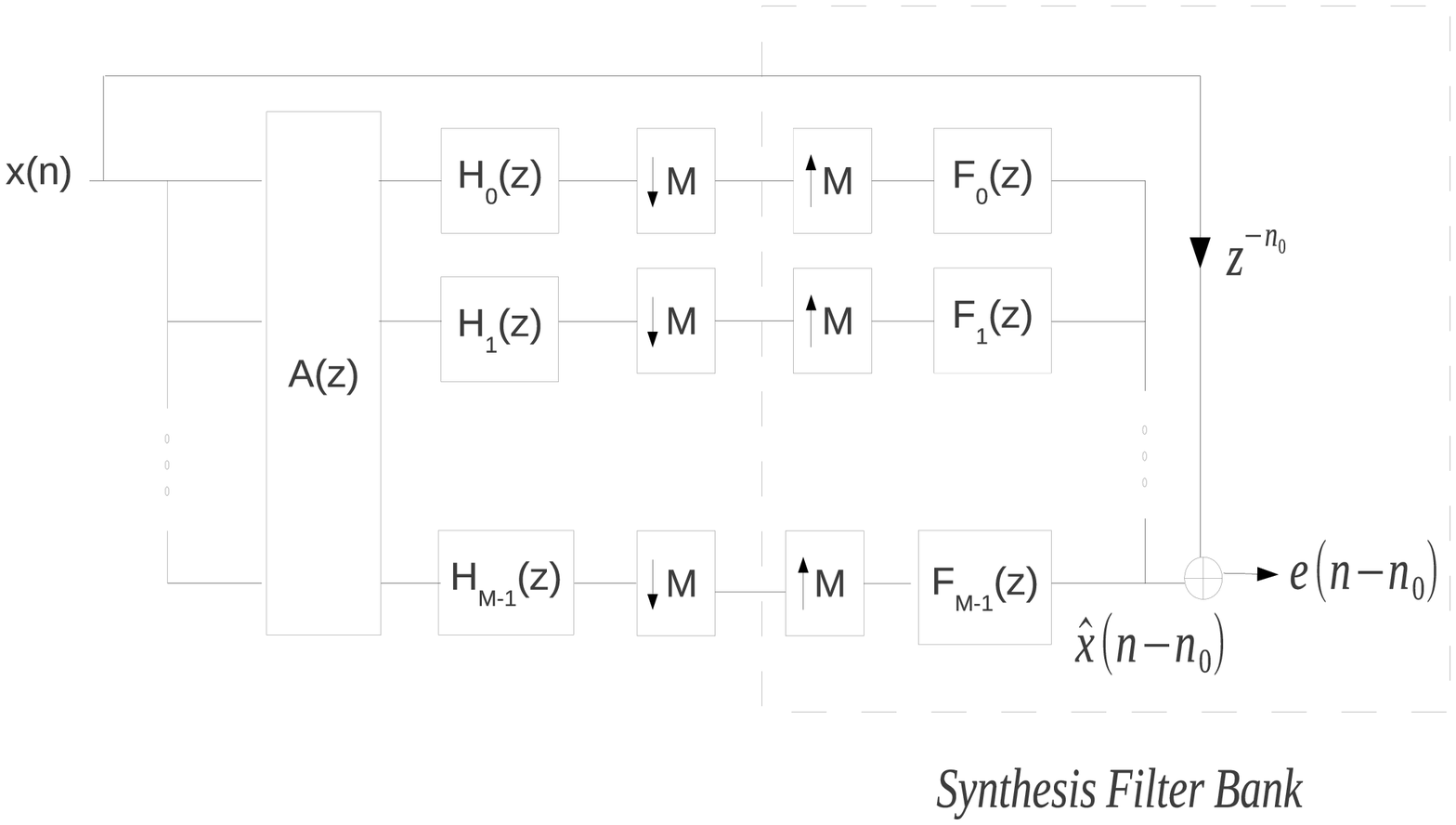}
\vspace{-80pt}
\caption{ Synthesis filter bank} \label{fig:msmfb}
\end{figure}

 In this paper, we define a concept of signal matched synthesis filter bank. The multi input single output multirate synthesis filter bank is first viewed as single input single output system. The signal matched synthesis filter bank is then defined as a solution to a prediction error problem, when the signal to be predicted is a delayed version of the input signal, fig.(\ref{fig:msmfb}) (delay is M-1, where M is the number of channels, as given in \cite{vetterli}). It is observed that the proposed synthesis filter bank can also be used as a model for generation of a signal. 

 \vspace{-0.5cm}
 
  We present a time and order recursive least squares algorithm
for the proposed signal matched synthesis filter bank. The recursions, so obtained, can be put in a
form of a lattice-like structure which consists of a circular lattice and M- joint process estimators. The filter parameters are, however, not directly available from this structure. For the estimation of these parameters, an order recursive algorithm is also presented. Simulation results are presented to demonstrate that the proposed algorithm fulfills both the objectives, mentioned earlier.\\

This paper is organized as follows: in section II, we define a concept of signal
matched, multirate, synthesis filter bank using a single input single output representation.
  We propose a geometrical framework, required for the given data case scenario, and discuss formulation of the problem. This is followed by the development of the required least squares algorithm. It is then observed that the recursions, so obtained, can be put in the form of a circular lattice-like structure. Since the signal matched synthesis filter bank parameters are not available from the algorithm, we present a computationally efficient, order recursive algorithm to obtain the same, in section III. In section IV, simulation
results are presented to validate the theory, for both the objectives mentioned above. Conclusions are presented in section V.

\subsection{Notations Used}
All the notations are same as given in section I of {\cite{PartI}}.

\section{Signal Matched Synthesis Filter Bank}

In this section we first consider the synthesis side, of an M-band, uniformly decimated, multirate filter bank, shown in fig.(\ref{fig:msmfb}). By making use of type-II polyphase decomposition, express this synthesis filter bank as an M-input M-output system. Then, by serializing the input as well as the output, convert the expression into a single input single output form. We then define the concept of signal matched synthesis filter bank by regarding this single input single output expression as a prediction error problem. We start by discussing some preliminaries.

\subsection{Preliminaries}

Consider the i-th filter of the M-band synthesis filter bank, shown in fig.(\ref{fig:msmfb})and is denoted as $F_{i}(z)$. Assuming the filter to be FIR and filter order to be multiple of M, without any loss of generality, we can write :
\begin{eqnarray} 
&& \rm {F}_{i}(z)= \sum_{n=0}^{N-1}{f_{i}(n)z^{-n}}\nonumber\\
&& =\left(\sum_{p=0}^{(N/M-1)}{{f}_{i}(pM)z^{-Mp}}\right)
+z^{-1}\left(\sum_{p=0}^{(N/M-1)}{{f}_{i}(pM+1)z^{-Mp}}\right)
\nonumber\\ 
&&+ \cdots +\rm  z^{-M+1}\left(\sum_{p=0}^{(N/M-1)}{{f}_{i}(pM+M-1)z^{-Mp}}\right)\nonumber\\
\end{eqnarray}

The M-components, written in braces, are the type-II polyphase components, so the k-th component, for the i-th filter can be written as:
\begin{eqnarray}
\rm {F}_{ik}(z^M) \triangleq \sum_{p=0}^{(N/M-1)}{{f}_{i}(pM+M-1-k)z^{-Mp}}
\end{eqnarray}

\begin{figure}[H]
\hspace{-0.5cm}
\includegraphics[width=0.8\textwidth, height=0.4\textwidth]{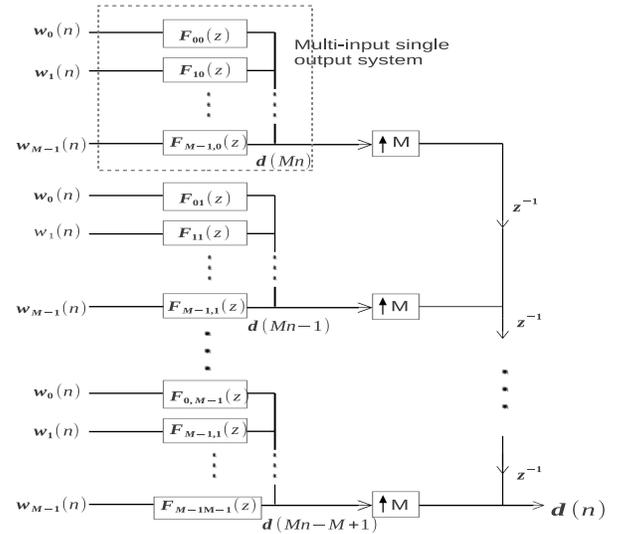}
\vspace{-10pt}
\caption{Polyphase decomposition of Synthesis filter bank} \label{fig:poly}
\end{figure}

 From the basic filter bank theory and the polyphase decomposition, as mentioned above, the synthesis filter bank can be re-structured into a multi input multi output system, shown in fig. \ref{fig:poly}. The outputs $d(Mn-i), 0 \leq i \leq M-1$, can be written as follows:


\begin{eqnarray}
{{\textit d}}(Mn-i)=\rm \sum_{j=0}^{M-1}{\sum_{p=0}^{N/M-1}{w_{j}(p)\textit{f}_{j}(Mn-i-Mp)}}\label{eq:poly}
\end{eqnarray}

The above set of equations, for $0 \leq i \leq M-1$, can be written in matrix form as follows:
\newcounter{tempfigcounter2}


\begin{footnotesize}
\begin{eqnarray} && \hspace*{-0.5cm}
\left[\begin{array}{c}
{d}(Mn-M+1)\\
{d}(Mn-M+2)\\
 \vdots \\
{d}(Mn)
\end{array}\right] =\left[\begin{array}{cccccc}
f_{0}(0) & f_{1}(0) & \ldots & f_{M-1}(0)\\
f_{0}(1) & f_{1}(1) & \ldots & f_{M-1}(1)\\
\vdots & \vdots & \ldots & \vdots\\
f_{0}(M-1) & f_{1}(0) & \ldots & f_{M-1}(M-1)
\end{array}\right]\nonumber\\&& \hspace*{3cm}
\left[\begin{array}{c}
\textit{w}_{0}(n)\\
\textit{w}_{1}(n)\\
\vdots \\
\textit{w}_{M-1}(n)
\end{array}\right]+...+
\nonumber \\&& \hspace*{1cm}
\left[\begin{array}{cccccc}
f_{0}(N-1) & f_{1}(N-1) & \ldots & f_{M-1}(N-1)\\
f_{0}(N-2) & f_{1}(N-2) & \ldots & f_{M-1}(N-2)\\
\vdots & \vdots & \ldots & \vdots \\
f_{0}(N-M) & f_{1}(N-M) & \ldots & f_{M-1}(N-M)
\end{array}\right]\nonumber \\&& \hspace*{3cm}
\left[\begin{array}{c}
\textit{w}_{0}(n-N+1)\\
\textit{w}_{1}(n-N+1)\\
\vdots \\
\textit{w}_{M-1}(n-N+1)
\end{array}\right]\label{eq:2}
\end{eqnarray}
\end{footnotesize}

 The multi input multi output system represented by(\ref{eq:2}) can be converted into a  single input single output form, shown in fig.(\ref{fig:poly_synth}), with the following two steps:

Step 1: Serializing the inputs:\\
 Pagano in \cite{pagano} 
  analyzed a multivariate AR processes using a periodic AR process, with fewer parameters. Later, Sakai \cite{sakai} used this relationship to present a Levinson-Durbin like algorithm for the multivariate AR process, the recursions in the algorithm lead to a circular lattice structure. In the same spirit, we define
%
  a scalar process $z$ using the input vector process, $\textbf{\textit{w}}(n)=\left[\begin{array}{cccc} w_{0}(n) & w_{1}(n) & \cdots w_{M-1}(n)
 
 \end{array}\right]$, as follows:

\begin{equation}
\textit{z}(Mn-i)=\textit{w}_i(n)
\end{equation} 

Step 2: Restructuring of filters:

 Also define filter $\rm G_{i}(z)=\sum_{n=0}^{N-1}{g_{i}(n)z^{-n}}$ for $0 \leq i \leq M-1 $, such that:
\begin{align}
\rm g_{(Mn-i) \_mod \_ (M)}(Mn-i+j)=f_{j}(Mn-i)
\end{align}

Substituting the above values in (\ref{eq:poly}), we obtain:

\begin{eqnarray}
{\textit{d}}(Mn-i)
=\sum_{j=0}^{N-1}
 {\textit g_{i}(j)z(Mn-j)}
 \label{eq:1}
\end{eqnarray}

The above equation gives a single input multi output representation, for $0 \leq M-1$, which can be converted into a single input single output system using the following step.

Step 3: Serializing the output:

Using M-fold upsamplers and a delay line, the M outputs of (\ref{eq:1}) are converted to a single output, $d(n)$, as shown in fig(\ref{fig:poly}).

\begin{figure}[H]
\vspace*{-1.5cm}
\hspace*{-1cm}
\includegraphics[width=0.6\textwidth, height=0.38\textwidth]{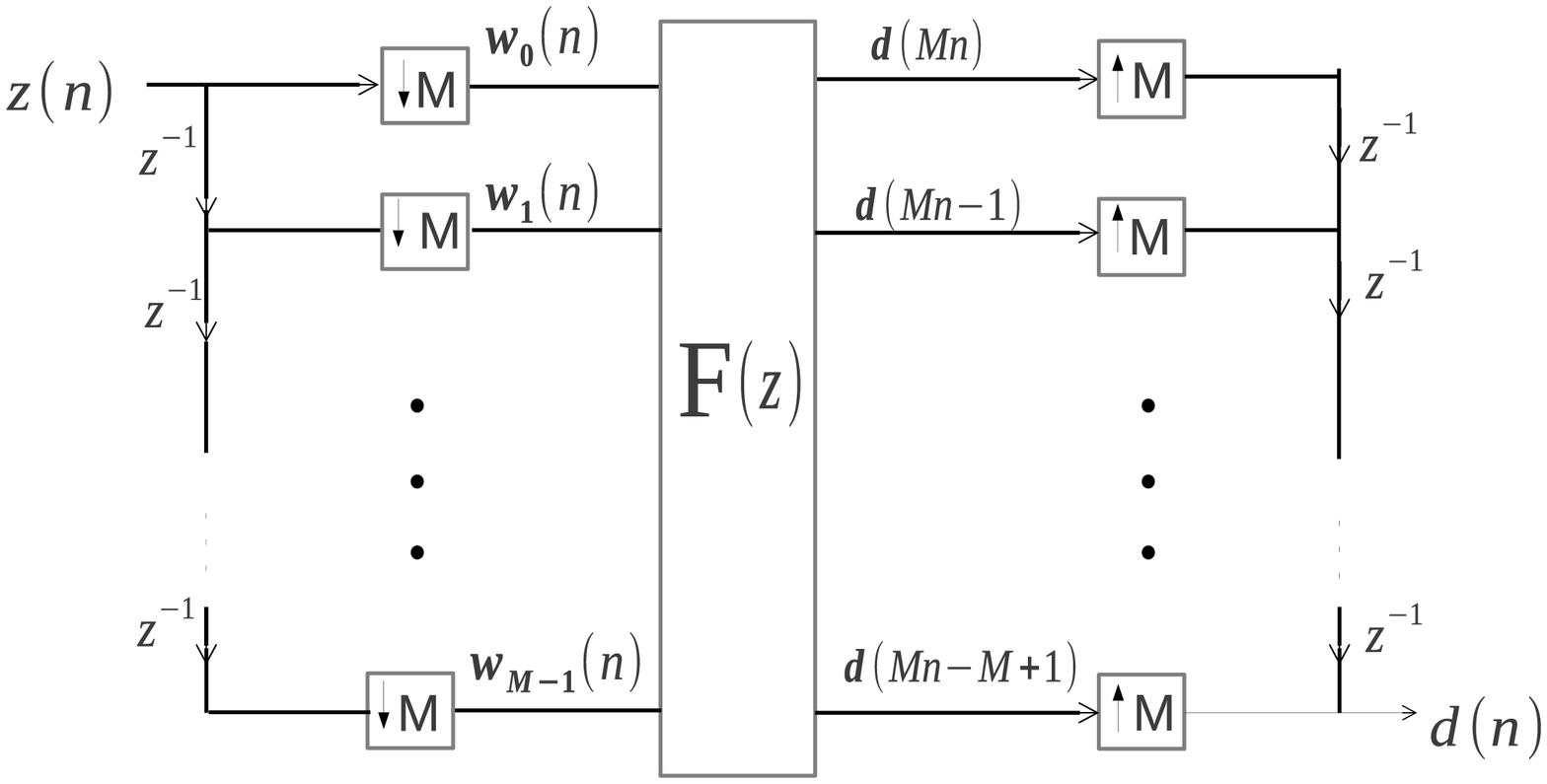}
\vspace{-60pt}
\caption{Single Input Single Output relationship of Signal Matched Synthesis Filter Bank} \label{fig:poly_synth}
\end{figure}
%
%
%
%
%
%

We now interpret (\ref{eq:1}) in two different ways:

(A). If we regard $z(n)$ to be a cyclostationary signal, with period M, then $w_{i}(n)$ for $0 \leq i \leq M-1$, could be considered as its stationary constituents. And hence (\ref{eq:1}) can be regarded as a multirate linear combiner estimating some desired signal $d(n)$. In particular if $z(n)$ is either stationary or cyclostationary white process, this system could be used as a generation model for $d(n)$.

(B). The second interpretation for (\ref{eq:1}) could be given as follows: Assuming we have a signal matched whitening filter bank for a signal $x(n)$. Hence, from part I of the sequel, we can get the whitened outputs, $w_{i}(n)$'s. Now (\ref{eq:1}) could be treated as an estimator (or predictor) for a delayed version of the input, say $x(n-n_0)$. The filter bank, designed to obtain minimum distance between $x(n-n_0)$ and its estimate has all the ingredients to be called signal matched synthesis filter bank.
\begin{align}
&e(Mn-i)=\textit{{d}}(Mn-i)-\sum_{j=0}^{N-1}{\textit {g({j})}\textit{{z}}(Mn-j)}\label{eq:8}
\end{align}
  We can now propose a precise definition for signal matched synthesis filter bank.

\subsection{Definition}

The single input single output prediction filter, represented by (\ref{eq:8}), is called a signal matched synthesis filter bank, if the signal to be predicted is a delayed version of the given signal and minimizes the distance between $x(n-n_0)$ and its estimate.


\subsection{Hilbert Space Framework}
 Equation (\ref{eq:8}), which define the concept of signal matched filter bank, plays the central role in the development of the least squares algorithm. In this section we present a Hilbert space framework, required to obtain the geometrical interpretation of (\ref{eq:8}), for the given data case scenario. The framework is then used to develop the required algorithm. We will follow the notations given in section IV of Part I. We make use of equation (\ref{eq:8}) and write the outputs of the i-th channel, for time upto Mn, with the pre-windowed assumption, in matrix form as follows:

\begin{small}\begin{eqnarray}
& \left[\begin{array}{ccccccc}
0 & \ldots & 0 & e^{N}(M-i) & e^{N}(2M-i) & \ldots & e^{N}(Mn-i)\\
\end{array}\right]
\nonumber\\
& =\left[\begin{array}{ccccccc}
0 & \ldots & 0 & d(M-i) & d(2M-i) & \ldots & d(Mn-i)\\
\end{array}\right]\nonumber\\
&-\left[
\begin{array}{cccc}
 g_{M+1-i}(0) & g_{M+1-i}(1) & \ldots & g_{M+1-i}(N-1)\\
 \end{array}\right]
 \nonumber\\
 &\left[
 \begin{array}{cccccc}
0 & \ldots & 0 & z(M) & \ldots & z(Mn)\\
0 & \ldots & 0  & z(M-1) & \ldots & z(Mn-1)\\
 \vdots\\
 0 & \ldots & 0  & z(M-N+1) & \ldots & z(Mn-N+1)
 \end{array}
\right]
\label{eq:9}
\end{eqnarray}\end{small}

\vspace{0.5cm}

A superscript N is added, to denote the order of the prediction error.
We can write the above equation using matrix notations, as given below:
\begin{eqnarray}
\mathbf{e}^{N}(Mn-i)= \mathbf{d}(Mn-i)-\mathbf{{g}}_{M+1-i}Z^{Mn}_{N}\label{eq:11}
\end{eqnarray}
where,\\
\hspace{-5pt}
$\mathbf {e}^{N}(Mn-i)=\left[\begin{array}{cccc} 0 & \cdots & 0 & {e}^{N}(M-i) \end{array}\right.$

\hspace*{3cm}$\left.\begin{array}{ccc} {e}^{N}(2M-i) & \cdots & {e}^{N}(Mn-i) \end{array}\right]$\\
 \hspace*{1cm}=$1 \times L$ error vector of the i-th branch, at time $Mn$ of order N,\\

$ \mathbf{d}(Mn-i)=\left[\begin{array}{ccccc} 0 & \cdots & 0 & {d}(M-i)& {d}(2M-i) \end{array}\right.$
 \hspace*{2cm}$\left.\begin{array}{cc}   \cdots & {d}(Mn-i)  \end{array}\right]$
\\
\hspace*{1cm}=$1 \times L$  desired data matrix at time $Mn$,\\

$\mathbf{z}(n)=\left[\begin{array}{cccc} 0 & \cdots & 0 & {z}(0) \end{array}\right.$
 $\left.\begin{array}{ccc} {z}(1) & \cdots & {z}(n)  \end{array}\right]$
\\ 
\hspace*{1cm}=$1 \times L$  input data matrix at time $Mn$, and\\

$\rm {Z^{Mn}_N}=\left[\begin{array}{cccc}\mathbf{z}(0)^T & \mathbf{z}(M)^T & \cdots & \mathbf{z}(Mn)^T  \end{array}\right]^T$
\\\hspace*{1cm}=$L \times N$ input data matrix at time Mn, $L \gg (n+1)$.


 

We define $\mathbf{e}^{N}(Mn-i)$  as the minimum norm error vector obtained by projecting $ \mathbf{d}(Mn-i)$ on the space spanned by $ \{ \mathbf{z}(Mn), \mathbf{z}(Mn-1), \cdots , \mathbf{z}(Mn-N+1) \} $.
%
%
%
%
Therefore, the least squares error can be written as:

\begin{eqnarray}
\mathbf{e}^{N}(Mn-i)=  \mathbf{d}(Mn-i)
{P^{\perp}}\left[\rm{Z^{Mn}_{N}}\right]\label{eq:13}
\end{eqnarray}
The above equation is the geometric interpretation of (\ref{eq:8}), for the given data case and will now be used to develop the time and order recursive least squares algorithm.

%
\subsection{Problem Statement}
With (\ref{eq:13}) at our disposal we can state the problem of this paper:
``given M-whitened output sequences of analysis filter bank, from Part I, $\{{w}_{i}(l),0\leq l\leq Mn,0\leq i\leq M-1\}$ at time Mn, obtain a fast time and order recursive algorithm to compute the error, in estimating $x(n-n_0)$, in the least squares sense, using the signal matched synthesis filter bank, as given in (\ref{eq:13}),
  and also compute time and order updates for the filter bank coefficients".

\subsection{Development of the algorithm}
With a well defined geometrical framework for signal matched synthesis filter bank as (\ref{eq:13}), discussed above, 
 we now develop a fast, 
time and order recursive, least squares algorithm. As stated in the problem, our objective is to compute $ \mathbf{e}(Mn-i)$ in an order and time recursive manner. For this purpose, we only need the latest value of the least squares error. Therefore post-multiply both the sides of 
(\ref{eq:13}) with $\mathbf{\pi}^{T}$ and also replacing fixed order N by a variable p we get:
\begin{eqnarray}
\mathbf{e}^{p}(Mn-i)=  \mathbf{d}(Mn-i)
{P^{\perp}}\left[\rm{Z^{Mn}_{p}}\right]\mathbf{\pi}^{T}\label{eq:14}
\end{eqnarray}

For the required least squares algorithm, recursions are computed by substituting proper values in the inner product update formula, as given in \cite{Friedlander}, for reference it is also given as (38) in  appendix of part I. For complete reconstruction of the algorithm, all the auxiliary quantities, arising in the development of the algorithm, are defined in table I and their correlations are given in table II. The substitutions are given in table III and the resulting algorithm is presented in table IV. This is a concise presentation of the algorithm, however, to get a understanding of how the recursions are obtained we present some sample calculations:

Example 1: Defining the auxiliary quantities:

 
 Substitute the following values in [(23),\cite{PartI}], first row of table I,
$\mathbf{y}=\mathbf{d}(Mn-M-i+1)$, $Y=Z_{p}^{Mn}$ and $\mathbf{w}=\mathbf{\pi}$, we obtain the definition of forward prediction error of i-th band:

\begin{align}
{e}^{p}(Mn-M+1-i)=\mathbf{d}(Mn-M-i+1)\rm P^{\perp}\left[Z_{p}^{Mn}\right]\mathbf{\pi}^{T}
\label{eq:r}
\end{align}

Example 2: Order update recursions:
 

 Substituting the following values in the inner product update formula given in \cite{Friedlander}, entries of first row of table III,  we obtain:\\
%
$\mathbf{\nu}=\mathbf{d}(Mn-M+1-i)$,
 $\rm{V_{1:n}}=\rm {Z_{p}^{Mn}}$,
  $\mathbf{w}=\mathbf{\pi}$ and
 $\mathbf{\nu_{n+1}}=\mathbf{z}(Mn)$

  \begin{eqnarray}
  e^{p+1}(Mn-i)=e^{p}(Mn-i)-\Delta_{e,\beta}^{p}(Mn-i).\nonumber\\
  R_{p}^{-\beta}(Mn).\beta^{p}(Mn)
  \end{eqnarray}
  


For the updates of the above equation, order updates of $\beta^{p}(Mn)$ is necessary, to obtain so, substitute the following values in the inner product update formula [(38),\cite{PartI}]:

$\mathbf{\nu}=\mathbf{z}(Mn-p-1)$,
 $\rm{V_{1:n}}=\rm{Z_{p}^{Mn-1}}$,
  $\mathbf{w}=\mathbf{\pi}$ and
 $\mathbf{\nu_{n+1}}=\mathbf{z}(Mn)$ 

 \begin{eqnarray}
 \beta^{p+1}(Mn)=\beta^{p}(Mn-1)-\Delta_{\beta,\alpha}^{p}(Mn-1).\nonumber\\
 R_{p}^{-\alpha}(Mn).\alpha^{p}(Mn)
 \end{eqnarray}
 

Similarly, all the recursions are obtained and put together in table \ref{table:LS_algo}. It can be easily seen that the above updates leads to a lattice structure, as shown in Fig.\ref{fig:circular_lattice}.

\renewcommand{\arraystretch}{1.2}
\newcounter{temptablecounter7}
\begin{figure*}
\setcounter{temptablecounter7}{\value{figure}}
\vspace{-20pt}
\begin{table}[H]
\begin{center}\begin{small}\caption{Definition of auxiliary quantities, for $0\leq i\leq M-1$ }\label{Table:aux_defined}
\begin{tabular}{|llllp{5cm}|}
\hline 
\multicolumn{1}{|l}{e} & \multicolumn{1}{l}{$\nu$} & \multicolumn{1}{l}{U} & \multicolumn{1}{l}{$w$} & comment
\tabularnewline
\hline 
$e^{p}(Mn-M+1-i)$   &     $\mathbf{d}(Mn-M+1-i)$  &  $\rm{Z^{Mn}_{p}}$ & $\pi$ & Forward prediction error of $i^{th}$ output band 
\tabularnewline
$\alpha^{p}(Mn-M+1-i)$ & $\mathbf{z}(Mn-M+1-i)$ & $\rm{Z^{Mn-M-i}_{p}}$ & $\mathbf{\pi}$ & Forward prediction error of $i^{th}$ band joint process estimator
\tabularnewline
$\beta^{p}(Mn-M-i)$ & $\mathbf{z}(Mn-M-i-p)$ & $\rm{Z^{Mn-M-i}_{p}}$ & $\mathbf{\pi}$ & Backward prediction error of $i^{th}$ band joint process estimator 
\tabularnewline
$\delta_{p}^{i}(Mn-M-i)$ & $\mathbf{\pi}$ & $\rm{Z^{Mn-M-i}_{p}}$ & $\mathbf{\pi}$ & $i^{th}$ band likelihood variable
 \tabularnewline
\hline 
\end{tabular}
\end{small}\end{center}
\end{table}
%
%
%
%
\begin{table}[H]
\begin{center}
\caption{Autocorrelation and cross-correlation coefficients}\label{table:corr}
\begin{tabular}{|lll|}
\hline 
$\mathbf{\nu}$ & $\mathbf{w}$ & $\mathbf{\nu w^{T}}$\tabularnewline
\hline 
$\mathbf{e^{p}}(Mn-M+1-i)$ 
 & $\mathbf{\beta^{p}}(Mn)$ & $\Delta_{e,\beta}^{p}(Mn-M+1-i)$\tabularnewline
$\mathbf{\alpha^{p}}(Mn-M+1-i)$ & $\mathbf{\alpha^{p}}(Mn-M+1-i)$ & $R_{p}^{\alpha}\left(Mn-M+1-i\right)$\tabularnewline
$\mathbf{\beta^{p}}(Mn-M-i)$ & $\mathbf{\beta^{p}}(Mn-M-i)$ & $R_{p}^{\beta}\left(Mn-M-i\right)$\tabularnewline
$\mathbf{\alpha^{p}}(Mn-M+1-i)$ & $\mathbf{\beta^{p}}(Mn-M-i)$ & $\Delta_{\alpha,\beta}^{p}(Mn-M+1-i)$\tabularnewline
$\mathbf{\beta^{p}}(Mn-M-i)$ & $\mathbf{\alpha^{p}}(Mn-M+1-i$ & $\Delta_{\beta,\alpha}^{p}(Mn-M-i)$\tabularnewline
\hline 
\end{tabular}\end{center}
\end{table}
%
%
%
\begin{table}[H]
\begin{center}\begin{small}
\caption{The substitution table}\label{table:substitution}
\begin{tabular}{|lllllll|}
\hline 
\multicolumn{1}{|l}{$v$} & \multicolumn{1}{l}{$\rm{V_{1:n}}$} & \multicolumn{1}{l}{$x$} & \multicolumn{1}{l}{$w$} & \multicolumn{1}{l}{$\nu_{n+1}$} & $v\rm{P^{\perp}}\left[x\right]\rm{P^{\perp}}\left[V_{1:n+1}\rm{P^{\perp}}\left[x\right]\right]w^{T}$ & \multicolumn{1}{l|}{}\tabularnewline
\hline 

$\mathbf{z}(Mn-M+1-i)$ & $\rm{Z^{Mn-M-i}_{p}}$ & $\mathbf{\pi}$ & $\mathbf{z}(Mn-M-i-p)$ & 0 & $\Delta_{\alpha,\beta}^{p}(Mn-M+1-i)$ & (1)\tabularnewline
$\mathbf{z}(Mn-M-i-p)$ & $\rm{Z^{Mn-M-i}_{p}}$ & $\mathbf{\pi}$ & $\mathbf{z}(Mn-M+1-i)$ & 0 & $\Delta_{\beta,\alpha}^{p}(Mn-M-i)$ & (2)\tabularnewline
 
$\mathbf{z}(Mn-M-i-p)$ & $\rm{Z^{Mn-M-i}_{p}}$ & $\mathbf{\pi}$ & $\mathbf{z}(Mn-M-i-p)$ & 0 & $R_{p}^{\beta}\left(Mn-M-i-p\right)$ & (3)\tabularnewline
 
$\mathbf{z}(Mn-M+1-i)$ & $\rm{Z^{Mn-M-i}_{p}}$ & $\mathbf{\pi}$ & $\mathbf{z}(Mn-M+1-i)$ & 0 & $R_{p}^{\alpha}\left(Mn-M+1-i\right)$ & (4)\tabularnewline

$\mathbf{\pi}$ & $\rm{Z^{Mn-M-i}_{p}}$ & 0 & $\mathbf{z}(Mn-M+1-i)$ & $\mathbf{\pi}$ & $\delta_{p+1}^{-1}(Mn-M-i)$ & (5)\tabularnewline
$\mathbf{z}(Mn-M+1-i)$ & $\rm{Z^{Mn-M-i}_{p}}$ & 0 & $\mathbf{\pi}$ & $\mathbf{z}(Mn-M-i-p)$ & $\alpha^{p+1}(Mn-M+1-i)$ & (6)\tabularnewline

$\mathbf{z}(Mn-M-i-p)$ & $\rm{Z^{Mn-M-i}_{p}}$ & 0 & $\mathbf{\pi}$ & $\mathbf{z}(Mn-M-i+1)$ & $\beta^{p+1}(Mn-M-i)$ & (7)\tabularnewline

$\mathbf{e^{p}}(Mn;-M+1-i)$ & 
$\rm{Z^{Mn}_{p}}$& $\mathbf{\pi}$ & $\mathbf{z}(Mn-p-1)$ & 0 & $\Delta_{e,\beta}^{p}(Mn-M+1-i)$ & (8)\tabularnewline
$\mathbf{x}(Mn-M+1-i)$ & 
$\rm{Z^{Mn}_{p}}$& 0 & $\mathbf{z}(Mn-p-1)$ & $\mathbf{\pi}$ & $e^{p}(Mn-M+1-i)$ & (9)\tabularnewline

\hline 
\end{tabular}
\end{small}\end{center}
\end{table}

\setcounter{figure}{\value{temptablecounter7}}
\end{figure*}
\renewcommand{\arraystretch}{1.6}
\newcounter{temptablecounter8}
\begin{figure*}
\setcounter{temptablecounter8}{\value{figure}}
\vspace{-20pt}
\begin{table}[H]
\begin {center}\begin{small}\caption{The LS algorithm for the synthesis side}\label{table:LS_algo}
\begin{tabular}{|ll|}
\hline 
$\begin{aligned} & Set\, all\,\Delta s\, and\, Rs\, to\,0\, at\, k=0\\
 & e^{0}(Mk-M+1-i)=d(Mk-M+1-i)\,`\,\, \alpha^{0}(Mk-M+1-i)=\beta^{0}(Mk-M+1-i)=d(Mk-M+1-i)\\
 & For\, p=1\, to\, N,\, i=0\, to\, M-1,k=0\, to\, n
\end{aligned}
$ & {} \tabularnewline
\hline 
$\Delta_{\alpha,\beta}^{p}(Mk-M+1-i)=\Delta_{\alpha,\beta}^{p}(M(k-1)-M+1-i))+\alpha^{p}(Mk-M+1-i)$.$\delta_{p}^{-1}(Mk-M-i).\beta^{p}(Mk-M-i)$ & (1)\tabularnewline
$\Delta_{\beta,\alpha}^{p}(Mk-M-i)=\Delta_{\beta,\alpha}^{p}(M(k-1)-M-i)+\beta^{p}(Mk-M-i).\delta_{p}^{-1}(Mk-M-i).\alpha^{p}(Mk-M+1-i)$ & (2)\tabularnewline
$R_{p}^{\beta}(Mk-M-i)=R_{p}^{\beta}(M(k-1)-M-i)+\beta_{p}(Mk-M-i).\delta_{p}^{-1}(Mk-M-i).\beta_{p}(Mk-M-i)$ & (3)\tabularnewline
$R_{p}^{\alpha}(Mk-M+1-i)=R_{p}^{\alpha}(M(k-1)-M+1-i)+\alpha^{p}(Mk-M+1-i).\delta_{p}^{-1}(Mk-M-i).\alpha^{p}(Mk-M+1-i)$ & (4)\tabularnewline
$\delta_{p+1}^{-1}(Mk-M-i)=\delta_{p}^{-1}(Mk-M-i)-\beta^{p}(Mk-M-i).R_{p}^{-\beta}(Mk-M-i).\beta^{p}(Mk-M-i)$ & (5)\tabularnewline
$\alpha^{p+1}(Mk-M+1-i)=\alpha^{p}(Mk-M+1-i)-\Delta_{\alpha,\beta}^{p}(Mk-M+1-i).R_{p}^{-\beta}(Mk-M-i).\beta^{p}(Mk-M-i)$ & (6)\tabularnewline
$\beta^{p+1}(Mk-M+1-i)=\beta^{p}((Mk-M-i)-\Delta_{\beta,\alpha}^{p}(Mk-M-i)R_{p}^{-\alpha}(Mk-M+1-i).\alpha^{p}(Mk-M+1-i)$ & (7)\tabularnewline
for j=0 to M-1 &{} \tabularnewline
$\Delta_{e,\beta}^{p}(Mk-M+1-i)=\Delta_{e,\beta}^{p}(M(k-1)-M+1-i)+e^{p}(Mk-M+1-i)
.\delta_{p}^{^{-1}}(Mk).\beta^{p}(Mk)$ & (8) \tabularnewline
$e^{p+1}(Mk-M+1-i)
=e^{p}(Mk-M+1-i)-\Delta_{e,\beta}^{p}(Mk-M+1-i).R_{p}^{-\beta}(Mk).\beta^{p}(Mk)
$ & (9)\tabularnewline
\hline 
\end{tabular}
\end{small}\end{center}
\end{table}
\setcounter{figure}{\value{temptablecounter8}}
\end{figure*}


\begin{figure*}
\begin{center}
\setcounter{tempfigcounter2}{4}
\begin{figure}[H]
\begin{center}
\vspace{-40pt}
\hspace{250pt}
\includegraphics[width=14cm,height=7cm]{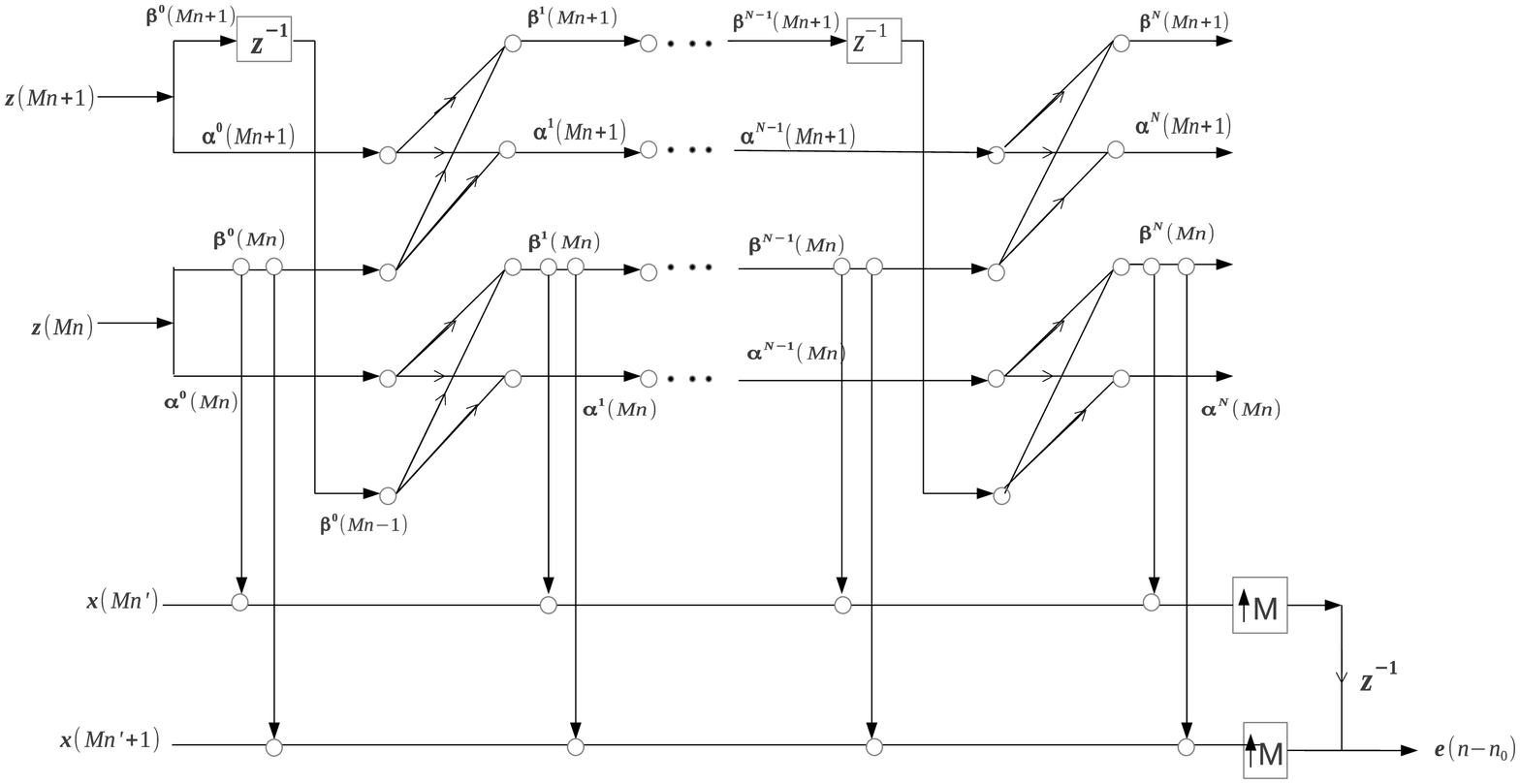}
\vspace{-40pt}
\caption{Lattice-Ladder structure for Synthesis filter bank} \label{fig:circular_lattice}
\end{center}
\end{figure}
\setcounter{figure}{\value{tempfigcounter2}}
\end{center}
\end{figure*}

\section{Least Squares Estimation of synthesis filter bank}
Since the proposed synthesis filter bank coefficients are not directly available from the algorithm, presented in the previous section, we will now propose a fast order and time recursive algorithm for estimation of these parameters, using the lattice coefficients. The recursions are computed in a similar manner as done in PartI\cite{PartI}, by proper substitutions in the pseudo-inverse update formula \cite{joshiII}. To concisely present the algorithm, the parameter vectors along with the auxiliary quantities are defined in table \ref{table:def_par_est}. The values to be substituted in the update relation are given in table {\ref{table:substitution_par_est} and the recursions obtained with these are collected in a tabular form , in table {\ref{table:par_est}}. A set of examples are presented here, to give an overview of algorithm.
%

Example: Definition of the auxiliary quantity:\\

In the update recursions of the required coefficient vector, some auxiliary quantities also arise. These are defined by substituting the entries of table \ref{table:def_par_est} in [(36)\cite{PartI}].
Substitute the following values in [(36) \cite{PartI}],\\
here,\begin{small} $\mathbf{e}=\mathbf{e_{i}^{p}}(n)$; $\mathbf{y}=\mathbf{z}$;
 ${Y}=\rm {Z_{p}^{Mn}}$
and $\mathbf{\theta}=\mathbf{{g}_{i}^{p}}$\end{small}, we get:

%
\begin{small}
\begin{eqnarray*}
\hspace{10pt}\mathbf{{g}_{i}^{p+1}}\hspace{-2pt}=\hspace{-2pt}-\mathbf{x}(Mn-i)\left[\rm{Z_{p+1}^{Mn}}\right]^{T}
\left[\rm {Z_{p+1}^{Mn}}\left[\rm{Z_{p+1}^{Mn}}\right]^{T}\right]^{-1}\label{eq:16}
 \end{eqnarray*}
\end{small}


Example: Update recursions:\\
To obtain the update recursions, substituting the following values in [(40),\cite{PartI}], for $p\leq M-1-i$:\\
$\mathbf{z}=\mathbf{x}(Mn-i)$;  $\rm V_{1:n-1}=\rm{Z_{p}^{Mn}}$; $x=0$ and\\$v_{n}=\mathbf{z}(Mn-p-1)$, we get:
\begin{eqnarray}
{\mathbf{{g}}}_{i}^{p+1} =\left[{\mathbf{{g}}}_{i}^{p} \mid 0\right]-
\Delta_{e_{i},\beta^{p}}(Mn-i)\nonumber\\
.R_{p}^{-\beta}(Mn).\left[\mathbf{d}^{p}(Mn-i) \mid 1\right]\label{eq:17}
\end{eqnarray}
where, $\Delta_{e_{i},\beta}^{p}(Mn-i)$ and $R_{p}^{-\beta}(Mn)$ are the coefficients of lattice ladder structure. From the above expression observe the relationship between the parameter vector of the filter bank and lattice coefficients. 
All the required recursions are collected in table \ref{table:par_est}.

\renewcommand{\arraystretch}{2}
\newcounter{temptablecounter4}

\subsection{Complexity}
The number of arithmetic operations required at every time step to compute the outputs using the least squares algorithm as given in table \ref{table:LS_algo} are: 9MN additions and 18MN multiplications and the computations required for parameter estimation, table \ref{table:par_est}, are $3MN^2$ additions and $3MN^2$ multiplications, where N is the order of the filters and M is the number of channels.

\begin{figure*}
\setcounter{temptablecounter4}{\value{figure}}
\begin{tiny}
\begin{table}[H]\begin{small}
\begin{center}
\caption{Definition of the auxiliary quantities, for $0\leq i\leq M-1,0\leq j\leq M-1$ }
\label{table:def_par_est}
\begin{tabular}{|llll|}
\hline 
\multicolumn{1}{|l}{e} & \multicolumn{1}{l}{$\nu$} & \multicolumn{1}{l}{U} & \multicolumn{1}{l|}{$\Theta$}\tabularnewline
\hline 
$\begin{aligned}\mathbf{e}^{p}(Mn+M-1-i)\end{aligned}
$ & $\mathbf{x}(Mn+M-1-i)$ & $\rm{Z_{p}^{Mn}}$ & $\mathbf{{g}_{i}^{p}}$\tabularnewline
$\mathbf{\alpha^{p}}(Mn+M-i)$ & $\mathbf{z}(Mn+M-i)$ & $\rm{Z_{p}^{Mn+M-1-i}}$ & $\mathbf{\widetilde{c}_{i}^{p}}$\tabularnewline
$\mathbf{\beta^{p}}(Mn+M-1-i)$ & $\mathbf{z}(Mn+M-i-p)$ & $\rm{Z_{p}^{Mn+M-1-i}}$ & $\mathbf{\widetilde{d}_{i}^{p}}$\tabularnewline
\hline 
\end{tabular}
\end{center}\end{small}
\end{table}\end{tiny}

\renewcommand{\arraystretch}{2}
\begin{tiny}
\begin{table}[H]\begin{small}
\begin{center}
\caption{The substitution table, for $0\leq i\leq M-1,0\leq j\leq M-1$}
\label{table:substitution_par_est}
\begin{tabular}{|llll|}
\hline 
\multicolumn{1}{|l}{$z$} & \multicolumn{1}{l}{$V_{1:n}$} & \multicolumn{1}{l}{$\nu_{n+1}$} &\multicolumn{1}{l|}{} \tabularnewline

\hline 
$\mathbf{x}(Mn+M-1-i)$ & $\rm{Z_{p}^{Mn-M}}$ &  $\mathbf{z}(Mn-M-p)$  & (3)\tabularnewline
 
$\mathbf{z}(Mn+M-1-i)$ & $\rm{Z^{Mn+M-2-i}_{p}}$ & $\mathbf{z}(Mn+M-1-i-p)$ & (1)\tabularnewline

$\mathbf{z}(Mn+M-1-i-p)$ & $\rm{Z^{Mn+M-2-i}_{p}}$ & $\mathbf{z}(Mn+M-1-i)$  & (2)\tabularnewline
\hline 
\end{tabular}
\end{center}\end{small}
\end{table}\end{tiny}

\renewcommand{\arraystretch}{2}
\begin{table}[H]\begin{small}
\begin {center}
\caption{The LS estimation of the synthesis bank parameters }
\label{table:par_est}
\begin{tabular}{|ll|}
\hline 
Set
$\begin{aligned} 
  {c_i^{0}}={d_i^{0}}=\mathbf{{g}_{i}^{0}}=1,k=0\, to\, K
\end{aligned}
$ & \tabularnewline
\hline 
$\mathbf{\widetilde{c}_{i}^{p+1}}(k)$$=\left[\mathbf{c_{p}}(k)\mid0\right]-\Delta_{\alpha,\beta}^{p}(k).R_{p}^{-\beta}(k-1).\left[\mathbf{\widetilde{d}_{i}^{p}(k-1)}\mid1\right]$ & (1)\tabularnewline
$\mathbf{\widetilde{d}_{i}^{p+1}}\left(k\right)$$=\left[\mathbf{0\mid d_{p}}(k-1)\right]-\Delta_{\beta,\alpha}^{p}(k-1)R_{p}^{-\alpha}(k).\left[\mathbf{1\mid\widetilde{c}_{i}^{p}(k)}\right]$ & (2) \tabularnewline
$\mathbf{{g}_{i}^{p+1}}(k)=\left[\mathbf{{g}_{i}^{p}}(k)\mid0\right]-\Delta_{e(Mn+M-1-i),\beta}^{p}(k).R_{p}^{-\beta}(k).\left[\mathbf{\widetilde{d}_{i}^{p}(k)}\mid1\right]$ & (3) \tabularnewline
\hline 
\end{tabular}
\end{center}\end{small}
\end{table}
\setcounter{figure}{\value{temptablecounter4}}
\end{figure*}

\section{Simulation Results}
As mentioned in the beginning of the paper, we had set forth two objectives for this paper. Now we present simulation results to demonstrate how these objectives have been attained. Whereas the section A deals with the first objective, i.e. design of the signal matched synthesis filter bank, corresponding to the signal matched whitening filter bank of Part-I, the simulation results in subsection B, are designed to corroborate the second objective of the paper. 

(A). We considered three type of filters: minimum phase, maximum phase and mixed phase, we then excite these filters with white noise which follow various types of distributions, here we are presenting results for three distributions:\\
Excitation 1: Gaussian distributed with zero mean and variance = 1,\\
Excitation 2: exponential distributed with mean = 1.5 and \\
Excitation 3: uniform distributed in the interval [-1,1]  \\
and the three filters are:\\
Filter 1:A minimum phase filter $H_1(z)=[1]/[z^2-0.6z+0.3600]$\\
Filter 2: A mixed phase filter $H_2(z)=[z^2-2.95z+1.90]/[z^3-1.30z^2+1.05z-0.325]$\\
Filter 3: A maximum phase filter  $H_1(z)=[z-1.4]/[z^2-0.6z+0.3600]$\\
The filter bank  structure, i.e. both analysis and synthesis filter banks (as proposed in this and the accompanying paper) is applied to these different input signals 
and the results are shown in Fig.\ref{fig:time_response_gaussian} and Fig.\ref{fig:time_response_filter2}.\\


\begin{figure}[H]
    \begin{subfigure}[b]{0.3\textwidth}
  \includegraphics[width=1.4\textwidth, height=0.5\textwidth]{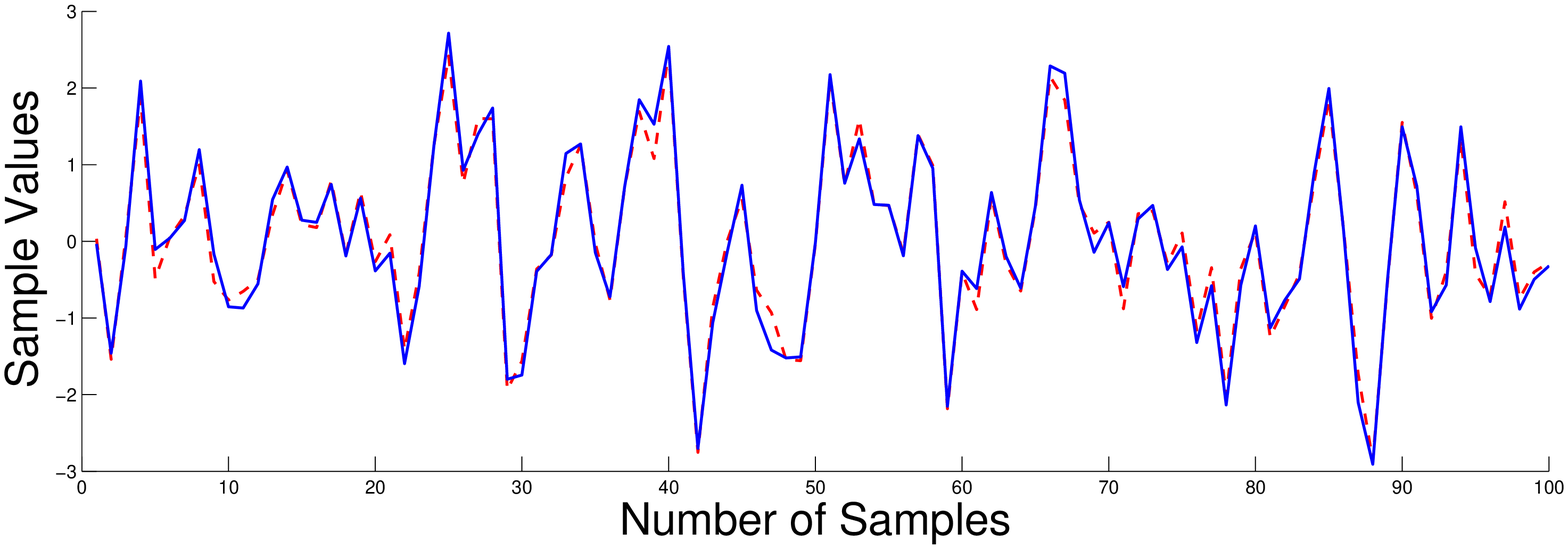}
        \caption{}
    \end{subfigure}
    
    \begin{subfigure}[b]{0.3\textwidth}
       
       \includegraphics[width=1.4\textwidth, height=0.5\textwidth]{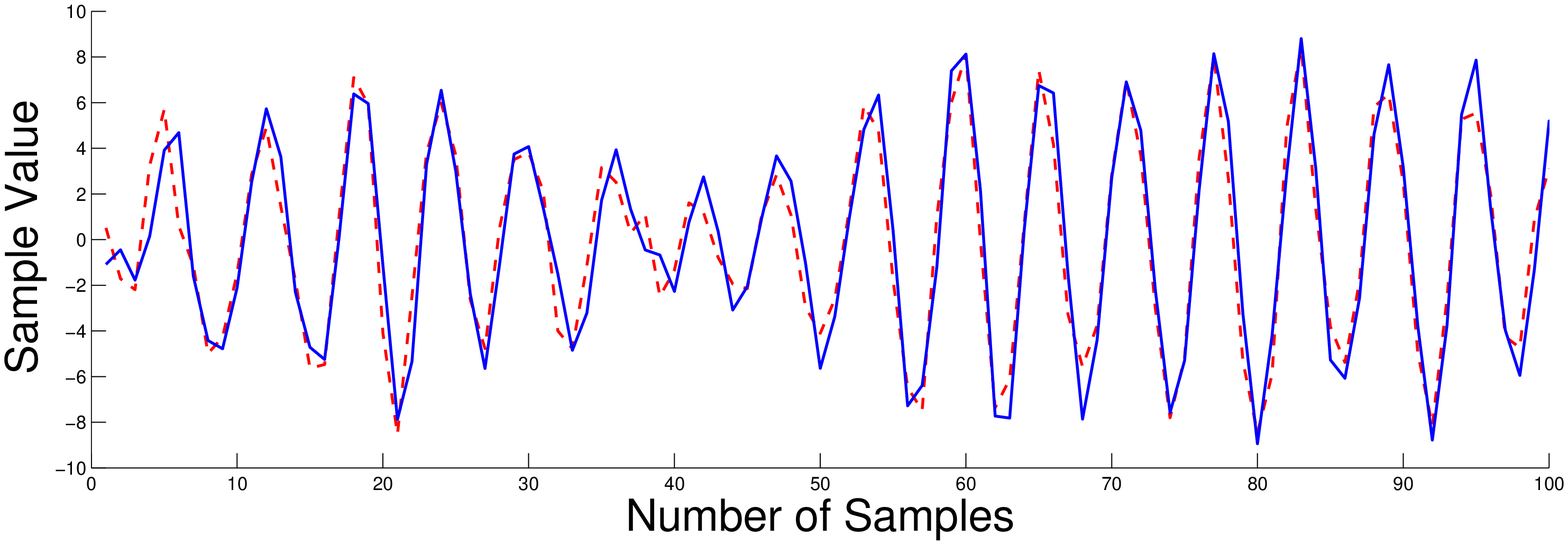}
        \caption{}
    \end{subfigure}
    \begin{subfigure}[b]{0.3\textwidth}
     
        \includegraphics[width=1.4\textwidth, height=0.5\textwidth]{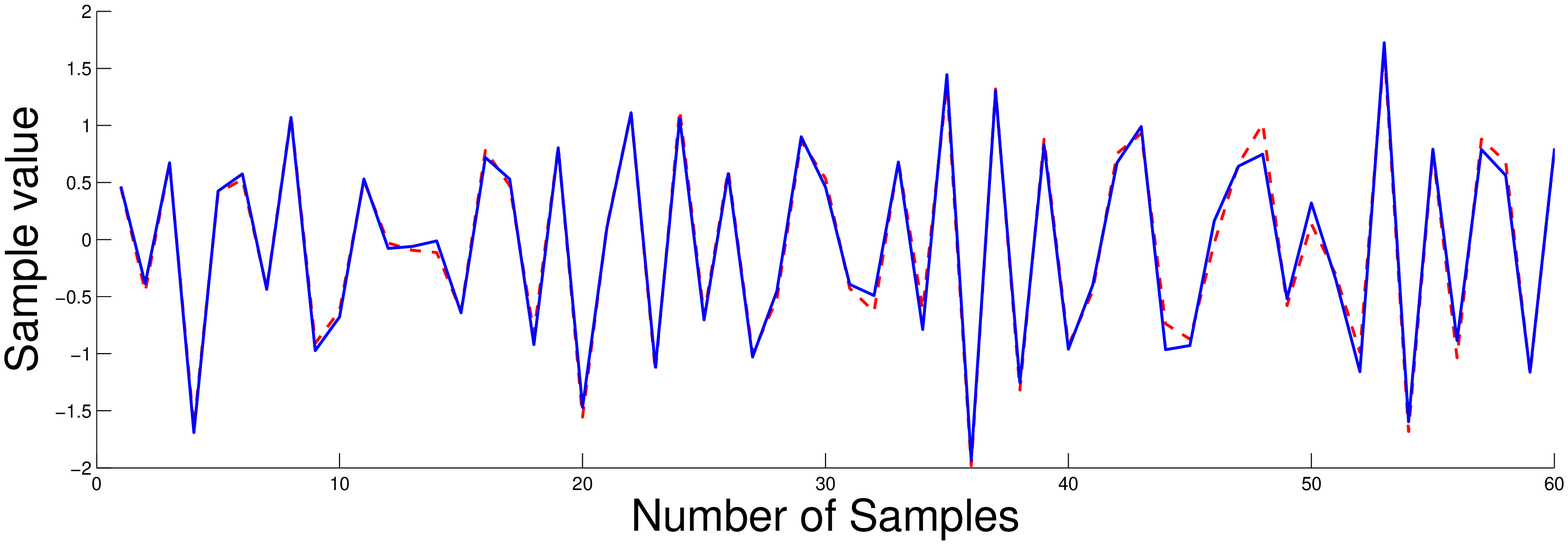}
        \caption{}
    \end{subfigure}
    \caption{Input signal, is represented by solid line, and estimated signal, is represented by dotted line, when input is :excitation 1 filtered with (a)Filter 1 (b) Filter 2 and (c)Filter 3.}
    \label{fig:time_response_gaussian}
\end{figure}

 \begin{figure}[H]
    \begin{subfigure}[b]{0.5\textwidth}
      \includegraphics[width=1\textwidth, height=0.3\textwidth]{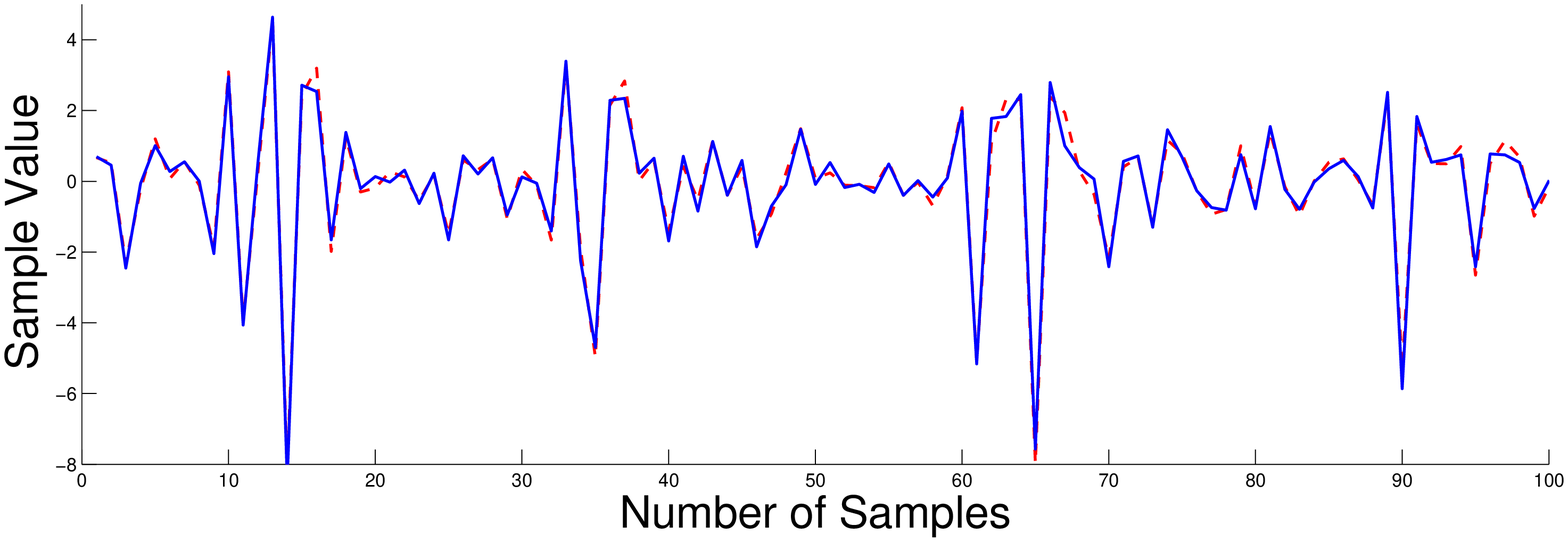}
        \caption{}
    \end{subfigure}
    \begin{subfigure}[b]{0.5\textwidth}
       \includegraphics[width=1\textwidth, height=0.3\textwidth]{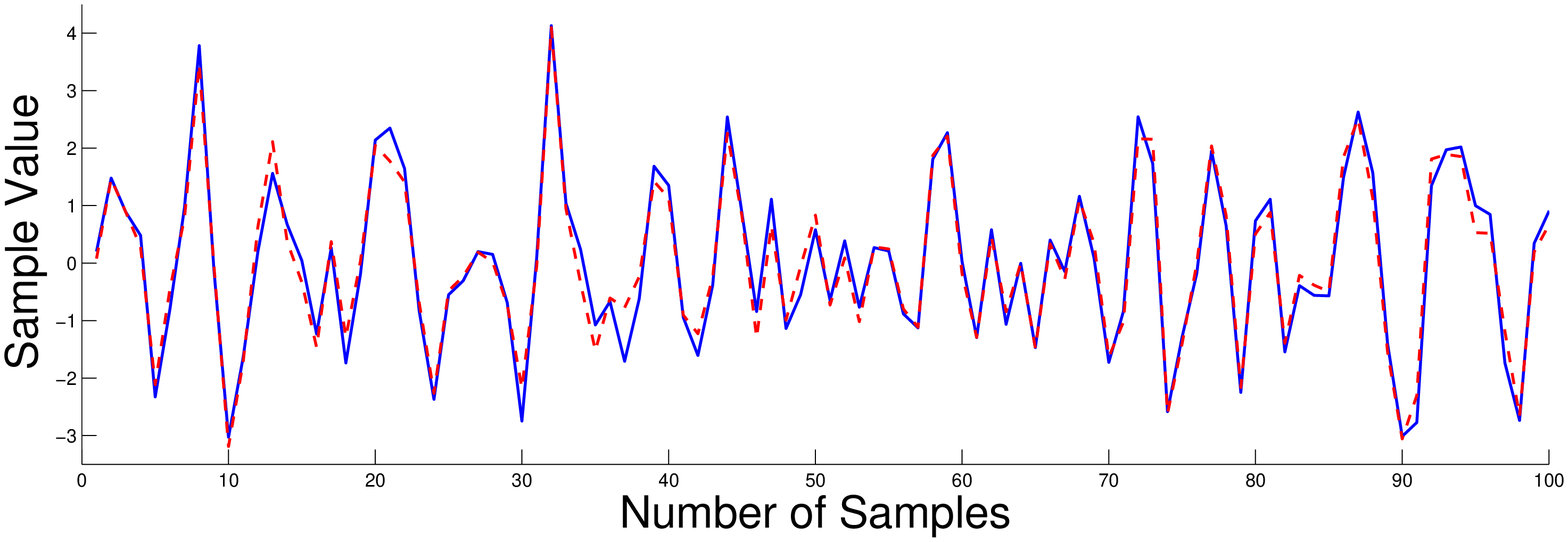}
        \caption{}
    \end{subfigure}
    
    \caption{Input signal, is represented by solid line, and estimated signal, is represented by dotted line, when the input signal is filtered using $H_{2}(z)$ and excited with white signal having (a) exponential distribution and (b) uniform distribution.}
    \label{fig:time_response_filter2}
\end{figure}

(B)Second objective of this work is to develop a modeling strategy for given signal. We have assumed that the given signal can be generated by passing M-white stationary processes through a uniform M-channel synthesis FIR filter bank. Thus to obtain the generation model of the given signal find the synthesis filter bank coefficients and obtain the required M-white processes using the the corresponding signal matched analysis side as proposed in Part I. To present simulation results for the second objective of the paper, we generate a synthetic signal using two WSS processes and a 2-channel uniform FIR synthesis filter bank of order 4, the coefficients of these filters are given in the first column of table \ref{Table:sim}. Now, using the proposed algorithm for signal matched filter bank \cite{PartI}, we obtain the required white processes. Further, the least squares algorithm presented in this paper are used to compute synthesis filter bank coefficients that will generate the given signal, these coefficients are given in the second column of table \ref{Table:sim}, these are recorded after the values converged.

\renewcommand{\arraystretch}{1.8}
\begin{figure*}
\setcounter{temptablecounter4}{\value{figure}}

\begin{table}[H]
\begin{center}\begin{small}\caption{Signal matched synthesis filter bank actual parameters and estimated parameters, for M=2 }
\label{Table:sim}
\begin{tabular}{|llll|}
\hline 
\multicolumn{1}{|l}{Signal Distribution} & \multicolumn{1}{l}{Filter} & \multicolumn{1}{l}{Actual parameters} & \multicolumn{1}{l|}{estimated parametrs} 
\tabularnewline
\hline
Gaussian distribution & $f_1(n)$ & $
[\begin{array}{cccc}-0.0167 &   0.0093 &
  -0.9976 & -0.6617\end{array} 
]

 $&
 $
  [\begin{array}{cccc} -0.0181 & 0.0113 &  -1.0483 & -0.6794 \end{array} 
] 
 $
\tabularnewline
& $f_2(n)$& $
[\begin{array}{cccc} -0.0179  & -0.9833 & -0.6267 & -0.0616\end{array} ]
$ & $

[\begin{array}{cccc} -0.0112 &  -1.0131  & -0.5975 & -0.0539 \end{array}] 
$
\tabularnewline
\hline 

Exponential Distribution & $f_1(n)$ & $
[\begin{array}{cccc}-0.1011 &   -1.0608 &
  -0.9633 & -0.5644 \end{array} 
]

 $&$
[\begin{array}{cccc} -0.1284  & -1.0547&  -0.9867 & -0.5568\end{array} ]
$
\tabularnewline
& $f_2(n)$& 
 $
  [\begin{array}{cccc} -1.0207 & -0.7548 &  -0.4785 & -0.4212 \end{array} 
] 
 $ & $

[\begin{array}{cccc} -1.0152 &  -0.7609  & -0.4162 & -0.3982 \end{array}] 
$
\tabularnewline
\hline

Uniform Distribution & $f_1(n)$ & $
[\begin{array}{cccc} 0.0344 &   -0.9589 &
  -0.5574 & -0.1044 \end{array} 
]

 $&
  $
[\begin{array}{cccc} 0.0293  & -0.9672 &  -0.5445 & -0.1169\end{array} ]
$
 
\tabularnewline
& $f_2(n)$& $
  [\begin{array}{cccc} -0.9910 & -0.5931 &  -0.0357 & 0.2547 \end{array} 
] 
 $& $\begin{aligned}
\left[\begin{array}{cccc} -0.9885 &  -0.5441&  -0.0481 & 0.2235 \end{array}\right] 
\end{aligned}$
\tabularnewline
\hline  

\end{tabular} 
\end{small}\end{center}
\end{table}
\setcounter{figure}{\value{temptablecounter4}}
\end{figure*}

\section{Conclusions}
A concept of signal matched synthesis filter bank is defined in this work, which perfectly reconstructs only the given signal and not every signal belonging to $L^2(\mathbb{R})$. The proposed signal matched synthesis filter bank when follows the analysis side as defined in \cite{PartI} gives a complete signal matched filter bank. An order and time recursive least squares algorithm is also proposed for the same. The signal matched synthesis filter bank is also presented as a modeling strategy for a given stochastic processes. Filter bank parameters are not directly available from the above mentioned algorithm and therefore an order recursive least squares algorithm is proposed to estimate these parameters using lattice coefficients. The theory proposed in this paper is validated using simulations results. In fig.\ref{fig:time_response_gaussian} and fig.\ref{fig:time_response_filter2}, we show that the proposed algorithm reconstructs Gaussian and non-Gaussian signals with minimum as well as non-minimum phase and   table \ref{Table:sim} illustrates that the proposed modeling scheme can be used for Gaussian as well as non-Gaussian signals.  

\bibliographystyle{IEEEtran}
\bibliography{IEEEabrv,reference}

\end{document}